\documentclass[rnote]{aa}  

\usepackage{graphicx}
\usepackage{txfonts}
\begin{document}

   \title{Merger relics of cluster galaxies}


   \author{Sukyoung K. Yi\inst{1}, Jaehyun Lee\inst{1}, Intae Jung\inst{1}, Inchan Ji
          \inst{1},
          \and
          Yun-Kyeong Sheen\inst{2}
              }
   \institute{Department of Astronomy and Yonsei University Observatory, Yonsei University, Seoul 120-749, Republic of Korea \\
              \email{yi@yonsei.ac.kr}
         \and
             Departamento de Astronom\'ia, Universidad de Concepci\'on, Casilla 160-C, Concepci\'on, Chile}

   \date{Received 27 February 2013; Accepted 16 May 2013}

 
  \abstract
   {Sheen and collaborators recently found that a surprisingly large portion (38\%) of massive early-type galaxies in heavy clusters show strong merger-related disturbed features. This contradicts the general understanding that massive clusters are hostile environments for galaxy mergers. Considering the significance of mergers in galaxy evolution, it is important to understand this.}
   {We aim to present a theoretical foundation that explains galaxy mergers in massive clusters.}
   {We used the N-body simulation technique to perform a cosmological-volume simulation and derive dark-halo merger trees. Then, we used the semi-analytic modeling technique to populate each halo with galaxies.  We ran hydrodynamic simulations of galaxy mergers to estimate the lifetime of merger features for the imaging condition used by Sheen and collaborators. We applied this merger feature lifetime to our semi-analytic models. Finally, we counted the massive early-type galaxies in heavy model clusters that would show strong merger features.}
   {While there still are substantial uncertainties, our preliminary results are remarkably close to the observed fraction of galaxies with merger features. Key ingredients for the success are twofold: firstly, the subhalo motion in dark haloes has been accurately traced, and, second, the lifetime of merger features has been properly estimated. As a result, merger features are expected to last very long in cluster environments. Many massive early-type galaxies in heavy clusters therefore show merger features {\em not} because they experience mergers in the current clusters in situ, but because they still carry their merger features from their previous halo environments. }
   {Investigating the merger relics of cluster galaxies is potentially important, because it uniquely allows us to backtrack the halo merger history.}

\titlerunning{Merger relics of cluster galaxies}
\authorrunning{Sukyoung K. Yi et al.}
   \maketitle

%

\section{Introduction}

A significant progress has recently been made toward a concordant picture of how our universe has achieved its current shape. One of the remaining big questions is the origin of massive galaxies. The current paradigm suggests that they are the result of numerous mergers between smaller galaxies \citep[e.g.,][]{whi78,col00}. When the masses of two colliding galaxies are similar such major mergers often result in a young starburst that easily outshines the underlying majority of old stellar populations \citep{law89} and sometimes even change the morphology of galaxies \citep[e.g.,][]{arp66,too72}. In this respect, understanding the frequency of mergers and their impact on galaxy evolution is the key to an accurate reconstruction of the apparently complex history of massive galaxy formation. 

Mergers between galaxies are expected to be much more frequent in low-density regions of the universe \citep{ost80}. This might appear counter-intuitive, but one can demonstrate this using advanced calculus \citep{bin87}. The essence of this prediction is that the spatial motion of galaxies is faster in a more massive halo, and mergers are less likely when relative speeds are higher. In these high-velocity environments, fly-by interactions are possible, but very unlikely to lead to galaxy mergers \citep{bin87}.

A challenge to this standard expectation was raised by the recent work of \citet{she12}. These authors used the Cerro-Tololo-Inter-American Observatory 4-meter Blanco Telescope to achieve $\mu_{\rm r}=28~$mag arcsec$^{-2}$ images of four massive galaxy clusters at $z \sim 0.1$. The unusually deep optical images revealed that a large portion (38\%) of red early-type galaxies showed strong merger features (tidal tails, shells, etc.). To everyone's surprise, the merger-featured galaxy fraction is almost similar to what was found earlier in low-density field environments, 49\% \citep{dok05}.

We propose to explain this enigma using the concept of merger relics. The theoretical expectation based on the merger probability as a function of halo size discussed above has an important caveat: cluster halos were assumed frozen while they are in reality alive. A cluster halo grows in size and mass with time as more galaxies fall into it. Furthermore, the galaxies falling in may have been part of another halo with multiple galaxies rather than being isolated. Therefore, it is important to first follow the halo merger history back in time and also determine whether the galaxies falling in have already completed their merging process that had started in the previous halo environment. This is a complicated task and calls for more elaborate calculations.

\section{Model construction}

A realistic reconstruction of the history of a cluster is challenging but possible using modern tools that are widely available. In the first step, we ran a cosmological-volume simulation with dark matter particles using the Gadget-2 code \citep{spr05}, assuming the WMAP-7 LCDM cosmology \citep{jar11}. The size of the cube was roughly 300 million lightyears on one side and the dark matter particle mass was $10^8~M_\odot$. Then we traced all virialized halos using the AdaptaHOP halo-finder program \citep{aub04,twe09}, and built halo merger trees. By doing this, we constructed full merger histories of all individual halos more massive than $10^{10}~M_\odot$. The basic information on the cosmological simulation runs is listed in Table 1.

Owing to poor mass resolution even in current simulations and technical challenges regarding halo-finding schemes, subhalos are often lost near the primary halo center where the background density can be higher than the densities inside the subhalos. For this reason, subhalos tagged as having merged sometimes reappear out of nothing in the halo outskirt \citep{oni12}. We attempted to fix this problem by keeping subhalos alive until they indeed fall into the host halo center. A short time before a subhalo was lost in the high-density region in the host halo, we started calculating its subsequent trajectory and lifetime using analytic formulae \citep{bin87,jia08}. This traces subhalos for longer than is possible with the halo finder we used and effectively lengthens the subhalo lifetime, roughly by a factor of two. The merging timescale between the host and subhalos in our simulation peaks around 1.3 billion years with a smooth lognormal distribution. The overall uncertainty in the merger rates derived this way is probably constrained within a factor of two \citep{hop10}.   
 
\begin{table}
\caption{Parameters of the cosmological simulation.}
\centering
\begin{tabular}{llc}
\hline\hline
Parameter	&	Description	&	Value	\\
\hline
$\Omega_{m}$	& Current matter density	&0.266	\\
$\Omega_{\Lambda}$		& Cosmological constant&0.734	\\
$\sigma_{8}$	& Initial power spectrum 	&0.801	\\
$n_{s}$	& Spectral index	&0.963	\\
$H_{0}$	& Hubble constant ($km s^{-1}Mpc^{-1}$)	&71.0	\\
\\
$L_{box}$	& Boxsize ($h^{-1}Mpc$)	&70	\\
$n_{part}$ & Number of particles	& $512^{3}$	\\
$m_{part}$ & Particle mass ($h^{-1}M_{\odot}$)	& $1.9\times 10^{8}$	\\
$\epsilon$ & Softening length ($h^{-1}kpc$) &$2.136$	\\
$n_{out}$ & Number of outputs	& 95 ($z=12$ to $z=0$)	\\

\hline
\end{tabular}
\end{table}

   \begin{figure}
   \centering
   \includegraphics[width=\hsize]{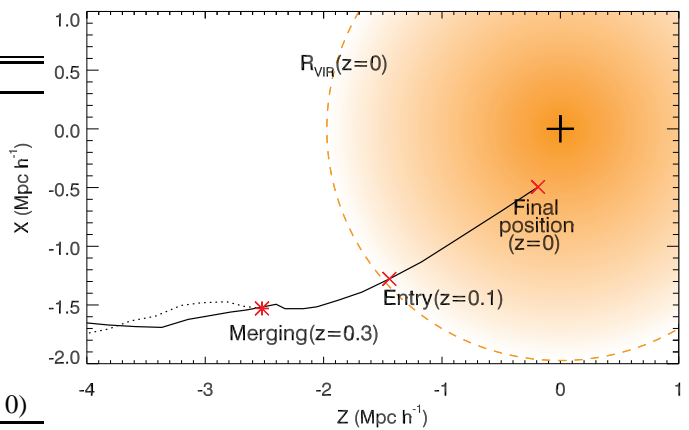}
      \caption{
      Example of the orbital evolution of a galaxy assumed to have merger features up to $z=0$ with respect to the center of a cluster in a comoving space. The solid and dotted lines show the tracks of the galaxy with a merger feature and its merger counterpart, respectively. The galaxy undergoes a major merger at $z=0.3$ and enters the virial radius of the cluster at $z=0.1$. The black cross indicates the center of the cluster and the orange dashed circle shows the virial radius of the host halo at $z=0$. The mass of the galaxy at $z=0.3$ is $1.4 \times 10^{10}~M_\odot$ and the mass of the merger counterpart is $1.0 \times 10^{10}~M_\odot$.
              }
         \label{f1}
   \end{figure}

In the second step, we constructed semi-analytic models of galaxies \citep[e.g.,][]{kau93,bau06,som08} based on the halo merger trees derived in the first step. These models allowed us to predict the number of galaxies of different types and properties that populate each halo. For most of the key ingredients, such as star formation, supernova, and active galactic nucleus feedback efficiencies, we used up-to-date parameters \citep{lee13}. For the environmental effects on the gas content in galaxies, we simply assumed that the distribution of hot gas follows a singular isothermal profile in halos. The hot gas components beyond a radius within which matter is considered to be gravitationally bound to subhalos are tidally stripped (see Kimm et al. 2011 for details). The amount of hot gas stripped by ram pressure was calculated in a similar manner. We computed a radius beyond which ram pressure can blow away hot gas components in subhalos into their host halos by adopting the prescriptions in McCarthy et al. (2008) and Font et al. (2008). Figure 1 shows an example of the orbital journey of a galaxy that encountered a major merger at $z=0.3$ and still shows its merger features at $z=0$.

Semi-analytic models provide a statistically large number of clusters and their constituent galaxies, but have no spatial information within each galaxy and hence are unsuited for investigating surface brightness profiles or tidal features. Hence an extra step is necessary. The third step is to estimate the lifetime of the merger features when a merger takes place. To quantify this, we performed hydrodynamic simulations of galaxy-galaxy mergers in an isolated environment; we used the GADGET-2 code with additional prescriptions including cooling of gas, star formation, and stellar feedback from type Ia and II supernovae \citep{pei09,pei10}. The galaxies in this study are modeled to represent the features (bulge-to-total ratio, gas mass fraction, and disk scale length) of Sa and Sb types. The total mass of each galaxy is $1.7 \times 10^{10}~M_\odot$. 

The distribution of gas and disk particles follows an exponential surface density while the distribution of dark matter (DM) and bulge particles follows the Hernquist profile (Hernquist 1990). We adopted the concentration parameter of $C = 14$ following Dolag et al. (2004) and the disk scale length of each galaxy from Graham \& Worley (2008). The particle masses are M(DM) = $4 \times 10^{5}~M_\odot$, M(disk) = M(gas) = $5 \times 10^{4}~M_\odot$, and M(bulge) = $1.5 \times 10^{5}~M_\odot$ for dark matter, disk, gas, and bulge, respectively. We used a gravitational softening length of 0.1 kpc for dark matter and bulge particles, and 0.2 kpc for disk and gas particles. The details of the galaxy properties used in the simulations are given in Table 2.

Radiative gas-cooling is based on the tabular function \citep{tho92,sut93}. We did not consider chemical evolution in the merger simulation but assumed all stars to have solar metallicity. For simplicity we explored only equal-mass mergers but considered various morphological combinations (between Sa and Sb types) and merger geometries (three orbital types and three merging angles). More detailed information on the simulations is provided elsewhere \citep{ji12} and will be available in the forthcoming companion paper (Ji, Peirani, \& Yi, in prep).

\begin{table}
\caption{Initial galaxy parameters for the merger simulations.}
\label{}
\begin{tabular}{lccc}
\hline\hline  &Description & Sa & Sb \\\hline
$M_{\rm vir}$ 					&Virial mass  $(M_{\odot})$	 	&$1.70 \times 10^{11}$	&$1.70 \times 10^{11}$\\$f_{\rm gas}$					&Disk mass fraction				&0.095				&0.107\\$f_{\rm bulge}$					&Bulge mass fraction			&0.063				&0.027\\$f_{\rm gas}$					&Gas mass fraction 				&0.008				&0.033\\B/T 							&Bulge-to-total-mass ratio	 		&0.4					&0.2\\$\lambda$   					&Disk spin parameter  			&0.05				&0.05\\		$R_{\rm d}$ 					&Disk scale length ${\rm (kpc)}$ 	&2.58				&3.29\\$b$ 							&Bulge scale length ${\rm (kpc)}$ 	&0.52				&0.66\\$z_{\rm 0}$ 					&Disk scale height ${\rm (kpc)}$ 	&0.52				&0.66\\$N_{\rm total}$   				&Total number of particles			&775,009			&857,875\\$N_{\rm halo}$   				&Number of halo particles			&353,142			&353142\\$N_{\rm disk}$ 					&Number of disk particles			&322,050			&361,600\\$N_{\rm bulge}$ 				&Number of bulge particles		&71,567			&30,133\\$N_{\rm gas}$     				&Number of gas particles			&28,250			&113,000\\
\hline

\end{tabular}\end{table}

The hydrodynamic simulation does not only trace old stars, but allows new star formation as well which has to be traced with particular caution. To perform ray-tracing, 2D regular Cartesian grids were constructed using the position of each particle. We convolved the stellar mass and age information of each stellar particle with the spectral energy distribution of \citet{bru03} to derive the light properties of model galaxies. To calculate the amount of dust extinction from hydrogen column density, we applied the empirical fitting formula from \citet{boh78} and the extinction formula of \citet{cal00} to the distribution of the gas and stellar particles in the simulations. The dust-to-gas ratios associated with metal abundances in their studies are close to those of Milky Way \citep{alt98,dav99}.
Then, we synthesized mock images by using the same filters as in the observations of Sheen et al. (2012).

Merger features in these synthetic images are spectacular and long-lasting. This is particularly dramatic when a faint limiting magnitude as achieved by Sheen et al. is used. Merger features are apparent soon after the first perigee pass and last until the end of our simulation run (4.5 billion years) in many cases (Figure 2). Merger features last roughly three times the coalescence time which is defined as the time required for the distance between the centers of the two galaxies to become zero. For a shallower surface brightness limit, the ratio of merger feature time to coalescence time, $t_{\rm m}/t_{\rm c}$, becomes lower. This is why merger features were difficult to detect in shallow surveys in the past. 

In reality, the lifetime of the merger features in the real universe is probably a more complex matter. Galaxy mergers occur between all morphological types and mass ratios, to begin with. The merging timescale is shortest for equal-mass mergers and thus our estimate for the merger feature time is likely a lower limit. On the other hand, mergers taking place in a cluster halo potential, as opposed to an isolated environment as in our simulation, may lose their merger features at different (probably shorter) timescales \citep{mih04}. 
The lifetime of merger features depends on the orbital geometry between merging galaxies, too. The features induced by merging on coplanar orbits were in general visible for longer (Ji 2012). The details will be presented in a separate paper (Ji, Peirani \& Yi, in prep.).
A full theoretical understanding will require an extensive simulation exercise filling a large parameter space. Until then, we chose the constraint from our calculations; that is, $t_{\rm m}/t_{\rm c} = 3$. Recalling that subhalos merge with host halos roughly at a timescale of 1.3Gyr in our cosmological volume simulation, the merger features are typically visible for 3.9Gyr given an image depth of $\mu_{\rm r}=28$mag arcsec$^{-2}$. This visibility time is calculated for each galaxy based on its unique merger history.

   \begin{figure}
   \centering
   \includegraphics[width=\hsize]{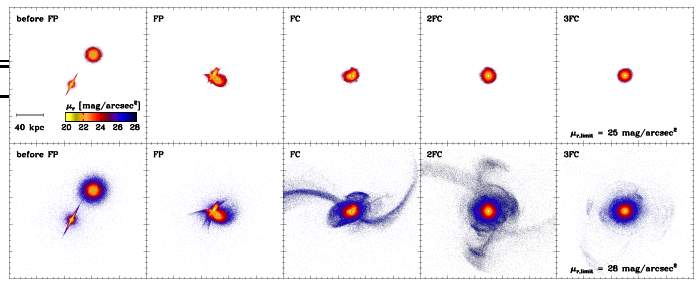}
      \caption{Mock images of an equal-mass merger between Sa and Sb galaxies, merging perpendicular to each other in a parabolic orbit (impact parameter 40 Kpc, pericentric distance 5 Kpc). Each row exhibits notable merger events in time sequence: i.e., before first perigee (FP) pass, first perigee pass, final coalescence (FC), 2 x FC, and 3 x FC, from left to right. The colour scheme represents surface brightness in the SDSS r' band. Top and bottom rows show the mock images in two different observing conditions, that is, $\mu_{\rm r}=25$mag arcsec$^{-2}$ (top) and $\mu_{\rm r}=28$mag arcsec$^{-2}$ (bottom). The deep mock images (bottom) correspond to the depth of the Sheen et al. observation.
              }
         \label{f2}
   \end{figure}

\section{Fraction of galaxies with merger features}

As a function of time, we counted bulge-dominant early-type galaxies that had a major merger in the last about 3.9 billion years, where we define major mergers as those with galaxy (stars+cold gas) mass ratio of 3 to 10 or higher. We examined only massive galaxies of stellar mass greater than $10^{10}~M_\odot$ to be consistent with the Sheen et al. observation. Galaxy morphology is classified based on a bulge-to-total stellar mass ratio, $B/T$. Galaxies with $B/T >0.4$ were classified as early-types  following \citet{som99}. Galaxy mergers play an important role in morphological evolution. In our model, mergers in general contribute to the growth or formation of bulges. Most notably, it is assumed that major mergers can dramatically change the galaxy morphology by disrupting disk structures. 

The end result is presented in Fig.~3. We inspected the 28 largest galaxy clusters in our simulation volume whose mass ranges M(halo) = $10^{14}$ through $5.6 \times 10^{14}~M_\odot$. Note in this figure that the fraction of massive early-type galaxies that are expected to show merger features solely from the merger relic process (ordinate) is remarkably high, roughly 35\% at $z=0.1$, surprisingly close to Sheen et al.'s observation. It was even higher in the past because major mergers were more frequent in the past when halos were smaller. The merger relic fraction is much higher for early-type galaxies than for late-type galaxies because we assumed that a falling galaxy merges into the central galaxy of the host halo, and the central galaxy is more likely to be early type.  

   \begin{figure}
   \centering
   \includegraphics[width=\hsize]{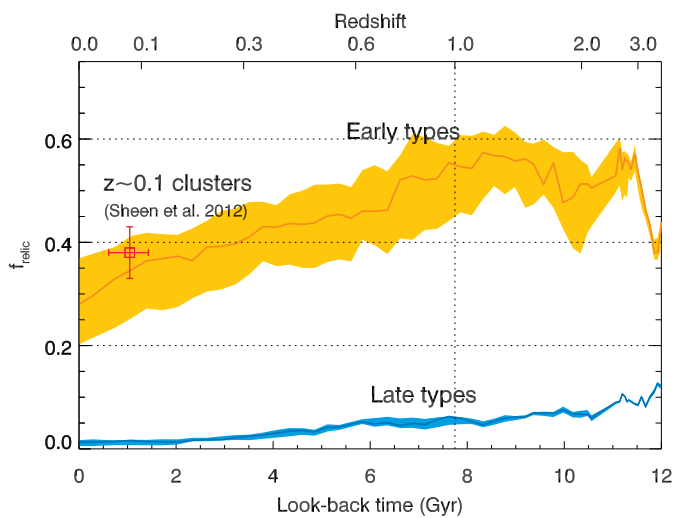}
      \caption{Evolution of the mean merger relic fraction based on 28 model clusters. A cosmological volume simulation has been run to build halo merger trees. Semi-analytic models of galaxy formation have been constructed based on the halo merger trees. Galaxy mergers were identified in this process. The lifetime of the merger features has been estimated based on hydrodynamic simulations of galaxy mergers, specifically for the target imaging depth of $\mu_{\rm r}=28$mag arcsec$^{-2}$. Then, we count massive early-type galaxies that still show their merger features since their latest major merger. Roughly 35\% of the massive early-type galaxies at $z \sim 0.1$ are expected to show post-merger features, which explains the recent finding of Sheen et al. The value has steadily declined since the early universe and is much higher for early types. Late-type (disk) galaxies in the clusters on the other hand are not expected to show major merger features so frequently.
              }
         \label{f3}
   \end{figure}

\section{Discussion and prospects}

While the level of agreement is good, there still are many sources of uncertainties. For one, it is subjective to find galaxies with major merger features, and so the merger fractions measured from observations and models are both uncertain, perhaps by a factor of two. Furthermore, this investigation does not include harassment due to the cluster potential. While harassment is thought to be more effective on small late-type galaxies \citep{moo96}, it may have some impact on massive early-type galaxies as well. Even considering all these uncertainties, it is probably safe to say that the merger relic fraction from our calculation ($\sim 35$\%) is much higher than the simple expectation based on the frozen halo scenario discussed in the beginning and surprisingly similar to what is observed.

We did not have to invent anything new for this success. Some ideas have been discussed in the past \citep{mih04,fuj04}, and we simply spliced pieces of modern and widely-used techniques to reach this explanation.  In-situ mergers may indeed be rare in heavy clusters where peculiar speeds are high. Yet, many massive early-type galaxies in heavy clusters exhibit strong merger features, which may be a lingering reflection from their past. 

We have learned from this exercise that correct interpretation of galaxy properties therefore requires a realistic understanding on their past environmental history in the first place. Conversely, investigating the properties and positions of merger-featured galaxies in a large number of clusters may even be used to reveal the past history of the clusters. On the numerical simulation side, a direct high-resolution simulation, probably adopting zoom-in technique, may soon replace this semi-analytic approach we present in this paper.

\begin{acknowledgements}
      SKY acknowledges support from National Research Foundation of Korea (Doyak Program No. 20090078756; SRC Program No. 2010-0027910) and DRC Grant of Korea Research Council of Fundamental Science and Technology (FY 2012). Numerical simulation was performed using the KISTI supercomputer under the program of KSC-2012-C2-11 and KSC-2012-C3-10. Much of this manuscript was written during the visit of SKY to University of Nottingham and University of Oxford under the general support by LG Yon-Am Foundation.
      
\end{acknowledgements}


\begin{thebibliography}{}
  \bibitem[Arp(1966)]{arp66} Arp, H. 1966, ApJS, 14, 1   
  \bibitem[Alton et al.(1998)]{alt98} Alton, P. B., Trewhella, M., Davies, J. I., Evans, R., Bianchi, S., Gear, W., Thronson, H., Valentijn, E., \& Witt, A. 1998, A\&A, 335, 807 
  \bibitem[Aubert et al.(2004)]{aub04} Aubert, D., Pichon, C., \& Colombi, S. 2004, MNRAS, 352, 376
  \bibitem[Baugh(2006)]{bau06}  Baugh, C. 2006, Reports on Progress in Physics, 69, 3101
  \bibitem[Binney \& Merrifiled(1998)]{bm98} Binney, J., \& Merrifiled, M. 1998, Galactic Astronomy (Princeton: Princeton Univ. Press)
  \bibitem[Binney \& Tremaine(1987)]{bin87} Binney, J., \& Tremaine, S. 1987, Galactic Dynamics (Princeton: Princeton Univ. Press)
  \bibitem[Bohlin, Savage \& Drake(1978)]{boh78} Bohlin, R. C., Savage, B. D., \& Drake, J. F. 1978, ApJ, 224, 132
  \bibitem[Bruzual \& Charlot(2003)]{bru03}  Bruzual, G., \& Charlot, S. 2003, MNRAS, 344, 1000
  \bibitem[Calzetti et al.(2000)]{cal00}  Calzetti, D., Armus, L., Bohlin, R. C., Kinney, A. L., Koornneef, J., \& Storchi-Bergmann, T. 2000, ApJ, 533, 682  
  \bibitem[Cole et al.(2000)]{col00} Cole, S., Lacey, C. G., Baugh, C. M., \& Frenk, C. S. 2000, MNRAS, 319, 168
  \bibitem[Cox et al.(2008)]{cox08}  Cox, T. J., Jonsson, P., Somerville, R. S., Primack, J. R., \& Dekel, A. 2008, MNRAS, 384, 386
  \bibitem[Davies et al.(1999)]{dav99}  Davies, J. I., Alton, P., Trewhella, M., Evans, R., \& Bianchi, S. 1999, MNRAS, 304, 495
  \bibitem[Dolag et al.(2004)]{dol04} Dolag, K., Bartelmann, M., Perrotta, F., Baccigalupi, C., Moscardini, L., Meneghetti, M., \& Tormen, G. 2004,   A\&A, 416, 853
  \bibitem[Font et al.(2009)]{fon09}  Font, A. S., Bower, R. G., McCarthy, I. G., et al. 2008, MNRAS, 389, 1619
  \bibitem[Fujita(2004)]{fuj04}  Fujita, Y. 2004, PASJ, 56, 29
  \bibitem[Graham \& Worly(2008)]{gra08} Graham, A. W. \& Worley, C. C. 2008, MNRAS, 388, 1708
  \bibitem[Hernquist(1990)]{her90} Hernquist, L. 1990, ApJ, 356, 359
  \bibitem[Hopkins et al.(2010)]{hop10}  Hopkins, P. F., Croton, D., Bundy, K. et al. 2010, ApJ, 724, 915
  \bibitem[Jarosik et al.(2011)]{jar11} Jarosik, N., Bennett, C. L., Dunkley, J. et al. 2011, ApJS, 192, 14
  \bibitem[Jiang et al.(2008)]{jia08}  Jiang, C. Y., Jing, Y. P., Faltenbacher, A., Lin, W. P., \& Li, C. 2008, ApJ, 675, 1095
  \bibitem[Ji(2012)]{ji12} Ji, I. 2012, M.S. Thesis (Yonsei Univeristy)  
  \bibitem[Kauffmann et al.(1993)]{kau93}  Kauffmann, G., White, S. D. M., \& Guiderdoni, B. 1993, MNRAS, 264, 201
  \bibitem[Kent, S. M., Dame, T. M. \& Fazio(1991)]{ken91} Kent, S. M., Dame, T. M., \& Fazio, G. 1991, ApJ, 378, 131
  \bibitem[Kimm et al.(2011)]{kim11}  Kimm, T., Yi, S. K., \& Khochfar, S. 2011, ApJ, 729, 1
  \bibitem[Lawrence et al.(1989)]{law89} Lawrence, A., Rowan-Robinson M., Leech, K., Jones, D. H. P., \& Wall, J. V. 1989, MNRAS, 240, 329
  \bibitem[Lee \& Yi(2013)]{lee13}  Lee, J. \& Yi, S. K. 2013, ApJ, 766, 38
  \bibitem[McCarthy et al.(2008)]{mcc08}  McCarthy, I. G., Frenk, C. S., Font, A. S. et al. 2008, MNRAS, 383, 593 
  \bibitem[Mihos(2004)]{mih04}  Mihos, J. C. 2004, in Clusters of Galaxies: Probes of Cosmological Structure and Galaxy Evolution, ed. J. S. Mulchaey, A. Dressler, \& A. Oemler (Cambridge: Cambridge Univ. Press), 277
  \bibitem[Moore et al.(1996)]{moo96}  Moore, B., Katz, N., Lake, G., Dressler, A., \& Oemler, A. 1996, Nature, 379, 613
  \bibitem[Onions et al.(2012)]{oni12} Onions, J. et al. 2012, MNRAS, 423, 1200 
  \bibitem[Ostriker(1980)]{ost80} Ostriker, J. P. 1980, Comments on Astrophys. 8, 177
  \bibitem[Peirani et al.(2009)]{pei09} Peirani, S., Hammer, F., Flores, H., Yang, Y., \& Athannassoula, E. 2009, A\&A, 496, 51
  \bibitem[Peirani et al.(2010)]{pei10} Peirani, S., Crockett, R. M., Geen, S., Khochfar, S., Kaviraj, S., \& Silk, J. 2010, MNRAS, 405, 2327  
  \bibitem[Sheen et al.(2012)]{she12} Sheen, Y., Yi, S. K., Ree, C. H., \& Lee, J. 2012, ApJS, 202, 8 
  \bibitem[Somerville et al.(2008)]{som08}  Somerville, R. S., Hopkins, P. F., Cox, T. J., Robertson, B. E., \& Hernquist, L. 2008, MNRAS, 391, 481
  \bibitem[Somerville \& Primack(1999)]{som99}  Somerville, R. S., \& Primack, J. R. 1999, MNRAS, 310, 1087
  \bibitem[Springel(2005)]{spr05} Springel, V. 2005, MNRAS, 364, 1105
  \bibitem[Sutherland \& Dopita(1993)]{sut93}  Sutherland, R. S. \& Dopita, M. A. 1993, ApJS, 88, 253
  \bibitem[Thomas \& Couchman(1992)]{tho92}  Thomas, P. A. \& Couchman, H. M. P. 1992, MNRAS, 257, 11
  \bibitem[Toomre \& Toomre(1972)]{too72} Toomre, A. \& Toomre, J. 1972, ApJ, 178, 623 
  \bibitem[Tweed et al.(2009)]{twe09}  Tweed, D., Devriendt, J., Blaizot, J., Colombi, S., \& Slyz, A. 2009, A\&A, 506, 647
  \bibitem[van Dokkum(2005)]{dok05} van Dokkum, P. 2005, AJ, 130, 2647 
  \bibitem[White \& Rees(1978)]{whi78} White, S. D. M., \& Rees, M. 1978, MNRAS, 183, 341 
\end{thebibliography}
\end{document}